\documentclass[english,10pt,pra,aps,showpacs,twocolumn,amsmath,amssymb,
longbibliography,
nofootinbib,
floats
]{revtex4-2}
\usepackage[T1]{fontenc}
\usepackage[latin9]{inputenc}
\usepackage{verbatim}
\usepackage{amsmath}
\usepackage{amssymb}
\usepackage{graphicx}
\usepackage{color}
\usepackage{bm}
\usepackage{hyperref}
\makeatletter

\usepackage{braket}
\begin{document}

\title{Confinement-induced resonance from the generalized Gross-Pitaevskii equations}

\author{Alexander Yu.~Cherny}
\email{cherny@theor.jinr.ru}
\affiliation{Bogoliubov Laboratory of Theoretical Physics, Joint Institute for Nuclear Research, 141980, Dubna, Moscow region, Russia}

\date{\today}

\begin{abstract}
The confinement-induced resonances for trapped bosons in the cigar-shaped and pancake geometries are studied within the generalized Gross-Pitaevskii equations, which are a simplified version of the Hartree-Fock-Bogoliubov approximation. Although the Hartree-Fock-Bogoliubov method is considered applicable only for small interparticle  interactions, the resonance denominators for the chemical potential are obtained in both quasi-one and quasi-two dimensions. A useful integral representation of the one-particle Green's function are found for the cylindrical confinement. We find the position of a smoothed resonance for the chemical potential in the pancake geometry at positive scattering length.
\end{abstract}

\maketitle

\section{Introduction}
\label{intro}

The confinement-induced resonance (CIR) is an efficient tool to tune effective interparticle interactions in low-dimensional systems of cold gases \cite{olshanii98,bergeman03,Petrov00,Petrov01,haller10} (see also review \cite{dunjko11}). The resonance was predicted in the cigar-shaped  \cite{olshanii98,bergeman03} and pancake \cite{Petrov00,Petrov01} geometries
and confirmed experimentally \cite{haller10}. The cases of very tight and anisotropic confinements were studied in Refs.~\cite{naidon07} and \cite{Peng10,Zhang11,melezhik11}, respectively.

The tight cylindrical or flat atomic waveguide creates, in effect, a low-dimensional system of cold atoms, and the problem is how to describe the corresponding interactions in low dimensions. The standard approach \cite{olshanii98,bergeman03,Petrov00,Petrov01} is to obtain interparticle interactions in low dimensions from the two-body scattering problem, thus reducing the effective dimension of the many-body system.

For identical bosons, it seems more attractive to develop a scheme that directly describes the wavefunction of the Bose-Einstein condensate  (BEC) for trapping potentials of arbitrary shape. This is all the more interesting because various optical traps are currently being produced experimentally, including an optical-box trap \cite{Lopes17,*Navon21}. The well-known Gross-Pitaevskii (GP) equation \cite{pitaevskii61,gross61,*gross63} is not suitabe for this purpose, since it does not reproduce the resonance denominators for effective interactions in low dimensions. For instance, in the cigar-shaped geometry, the wavefunction of BEC coincides up to a factor with the ground state of the trapping harmonic oscillator in the $x$-$y$ plane provided its frequency is sufficiently high. Then one can integrate out the $x$ and $y$ coordinates in the interaction term of the GP functional and obtain the effective coupling constant in one dimension \cite{Pitaevskii16:book}: $g_{\text{1D}}= {2\hbar^2a}/{(m\,l^2)}$. This result is valid only away from the resonance when $a\ll l$. Here $a$, $m$, and $l$ are the three-dimensional (3D) scattering length in free space, boson mass, and length of the two-dimensional (2D) oscillator, respectively. On the other hand, the correct coupling constant contains the resonance denominator \cite{olshanii98}, see Eq.~(\ref{eff_g1D}) below.

In the paper \cite{cherny04}, a system of non-linear equations in both stationary and time-dependent cases was proposed and called the generalized Gross-Pitaevskii (GGP) equations. Apart from the standard one-body wavefunction of BEC, the equations also contain the two-body wavefunction of BEC.  This approach is nothing else but a simplified version of the full Hartree-Fock-Bogoliubov (HFB) approximation (see Appendix \ref{sec:HFB} below). The GGP equations are applicable for arbitrary shape of the trapping potential. In particular, the resonance denominators were obtained \cite{cherny04} in both cases of cigar-shaped and pancake geometries. However, the resulting denominators of Ref.~\cite{cherny04}, which determine the position of the resonances, contained incorrect coefficients, since the solutions were derived with an additional quite rough approximation (see the discussion in Sec.~\ref{AppGF1} below). As is shown in this paper, the GGP equations indeed yield the exact coefficients, which are consistent with the two-body scattering problem \cite{olshanii98,bergeman03,Petrov00,Petrov01}.

The GGP equations are nonlinear, but the effective coupling constants in low dimensions are determined by the short-range correlations, where the nonlinearity is not important. The problem can be reduced to the two-body problem and the corresponding Green's functions (GFs). We find and examine the integral representations for the one-particle GFs at low energies for the flat and cylindrical harmonic confinements. In the latter case, the equation for the GF differs from the previously found integral representation \cite{Idziaszek06,*Idziaszek15,Liang08}, which is actually the Fourier transform of the Feynman propagator for a harmonic oscillator with contour rotation in the complex plane. We believe that the found representation is more convenient for obtaining the low-energy expansions of the GF.

This paper is organized as follows. In the next section, the GGP equations are written and discussed in the case of the tight confinements. In Sec.~\ref{sec:GF}, the integral representations and expansions of the GFs are obtained for the cylindrical and flat confinements. The CIRs for the effective coupling constant and chemical potential are found within the suggested scheme in Secs.~\ref{CIR1D} and \ref{CIR2D}. The main results and prospects are discussed in Conclusions. In Appendix \ref{sec:HFB}, the GGP are derived through the HFB approximation.

\section{Generalized Gross-Pitaevskii equations}
\label{sec:GGPeq}

We consider a dilute many-body system of identical bosons, interacting with a pairwise short-range\footnote{A potential is of the short-range type if it decreases at infinity as $1/r^\alpha$ with the exponent $\alpha > 3$ or faster (see, e.g., Ref.~\cite{jeszenszki18}). For such a potential, only the $s$-wave scattering amplitude survives in the limit of zero scattering energy. It is called the scattering length [see Eq.~(\ref{phi3dasymp}) below].}  potential $V(r)$, in an external field $V_\mathrm{ext}(\bm{r})$. For simplicity, the bosons are assumed to be spinless. The equilibrium properties of such systems at zero temperature are described by Eqs.~(16) and (17) of Ref.~\cite{cherny04}, which can be written in the form
\begin{align}
  &\big[\hat{H}(\bm{r}_{1})\! -\! \mu\big]\phi(\bm{r}_{1})\!+\!\int\! d\bm{r}_{2} \phi(\bm{r}_{2})V(|\bm{r}_{1}\!-\!\bm{r}_{2}|)\varphi(\bm{r}_{1},\bm{r}_{2})\!=\!0,   \label{1body}\\
  &\big[\hat{H}(\bm{r}_{1})\! +\! \hat{H}(\bm{r}_{2})\!-\!2\mu\big]\psi(\bm{r}_{1},\bm{r}_{2})\!+\!V(|\bm{r}_{1}\!-\!\bm{r}_{2}|)\varphi(\bm{r}_{1},\bm{r}_{2})\!=\!0, \label{2body}
\end{align}
where $\mu$ is the chemical potential. Here we set by definition
\begin{align}\label{psidef}
\psi(\bm{r}_{1},\bm{r}_{2})= \varphi(\bm{r}_{1},\bm{r}_{2}) - \phi(\bm{r}_{1})\phi(\bm{r}_{2}),
\end{align}
and the one-body Hamiltonian is given by
\begin{align}\label{1bodyHam}
\hat{H}(\bm{r})=-\frac{\hslash^2\nabla_{\bm{r}}^2}{2m} + V_\mathrm{ext}(\bm{r})-E_0
\end{align}
with $E_0$ being its ground-state energy. By definition, $\phi(\bm{r}) \equiv \langle \hat\Psi({\bm{r}})\rangle$ and $\varphi(\bm{r}_{1},\bm{r}_{2}) \equiv \langle \hat\Psi({\bm{r}_{1}})\hat\Psi({\bm{r}}_{2})\rangle$ are the anomalous averages of the field Bose operators. The derivation of these equations and their relation to the HFB approximation are given in Appendix \ref{sec:HFB}.

The function $\phi(\bm{r})$ is the one-body function of inhomogeneous BEC and can be considered as the order parameter. The function $\varphi(\bm{r}_{1},\bm{r}_{2})$ is called pair wavefunction. It describes two bosons in the Bose-Einstein condensate, and this function is \emph{not} reduced in general to the product of the one-body functions. The exact definitions can be done through the one-body and two-body matrices~\cite{cherny00pair,cherny00,cherny01}. The property of decay of correlations dictates the boundary condition
\begin{align}\label{bc}
 \varphi(\bm{r}_{1},\bm{r}_{2}) \simeq \phi(\bm{r}_{1})\phi(\bm{r}_{2})\quad \text{when} |\bm{r}_{1}-\bm{r}_{2}|\gg \xi_\mathrm{h},
\end{align}
where $\xi_\mathrm{h}$ is the healing (or coherence) length defined through the chemical potential: $\xi_\mathrm{h}=\hbar/\sqrt{\mu m}$. The system of equations (\ref{1body}) and (\ref{2body}) is nonlinear due to the presence of the condensate wavefunction $\phi(\bm{r}_{2})$ under the integral in Eq.~(\ref{1body}). The GGP equations (\ref{1body}) and (\ref{2body}), being a particular case of HFB approximation, are applicable to weak interactions when the mean distance between bosons is much smaller than the healing length.

Various solutions for different external potentials were considered in Ref.~\cite{cherny04}. In particular, the 1D and 2D harmonic CIRs were obtained [see Eqs.~(48) and (57) of Ref.~\cite{cherny04}]. However, the method used did not enable us to obtain the precise coefficients in the resonance denominators, which determine the CIR positions. In this paper, we show that the correct coefficients can be derived from Eqs.~(\ref{1body}) and (\ref{2body}).

We assume that the radius $r_{0}$ of the short-range interaction potential $V(r)$ is much smaller than the characteristic length of the external potential $V_\mathrm{ext}(\bm{r})$ and the mean distance between bosons. This implies that at low energies, physical processes are governed by the $s$-wave scattering length $a$ of the interaction potential, which is obtained from the symmetric solution of the two-body Scr\"odinger equation at zero energy in free space.

Thus, our task is to get the chemical potential of $N$ strongly confined bosons at zero temperature as a function of the 3D scattering length $a$ and oscillator frequency $\omega$ in the case of cylindrical and flat harmonic confinements.

%------------------------------------------------------------
\section{Green's functions of the one-body Schr\"odinger equation at low energies for cylindrical and flat harmonic confinements}
\label{sec:GF}

To obtain the chemical potentials of the Bose-Einstein condensates for the cigar-shaped and pancake geometries, we need to calculate the corresponding GFs for the one-body Hamiltonian (\ref{1bodyHam}), which is defined as
\begin{align}\label{GFdef}
\hat{G}_E=-(\hat{H}-E)^{-1}
\end{align}
with the scattering energy $E$, which is assumed to be sufficiently small (see the discussions below).

In this section, we find useful integral representations for the GFs in the cases of cylindrical and flat harmonic confinements. These representations are very convenient for obtaining the low-energy expansions of the GFs.

\subsection{Cylindrical confinement}
\label{sec:cyl}

The cylinder harmonic confinement implies that a particle is confined in the $x$-$y$ plane but can move freely in $z$-direction with a finite momentum, whose absolute value is denoted by $\hbar p$. The one-particle Hamiltonian (\ref{1bodyHam}) is given by
\begin{align}\label{1dconfham}
  \hat{H}=-\frac{\hslash^2\nabla^2}{2m_{0}} + \frac{m_{0}\omega^2(x^2+y^2)}{2}-E_0,
\end{align}
where the shift $E_0=\hbar\omega$ is introduced to get zero ground-state energy.

It is convenient to use the following notation for radius vector: $\bm{r}=(\bm{\rho},z)$ with $\bm{\rho}=(x,y)$. We calculate the GF (\ref{GFdef}) in coordinate representation with the radial coordinate $\rho=\sqrt{x^2+y^2}$ being equal to zero and $E=\frac{\hbar^2p^2}{2m_{0}}$:
\begin{align}
  &G_{p}(z-z')\equiv\langle z,\rho=0|\hat{G}_E|\rho'=0,z'\rangle\nonumber \\
  &=-\frac{1}{2\pi^2l^2}\sum_{n=0}^{\infty}\text{p.v.}\!\!\int_{-\infty}^{\infty}dq\frac{e^{iq(z-z')}}{2\hbar\omega n+\frac{\hbar^2}{2m_{0}}(q^2-p^2)}.
\label{GFcyl}
\end{align}
Here $l_{0}=\sqrt{\hbar/(m_{0}\omega)}$ is the harmonic oscillator length, and p.v. denotes the Cauchy principal value. This kind of regularization ensures the reality of the GF in coordinate representation \cite{cherny01a}.

Here we use the expression for the GF in Dirac notation $\langle \bm{r} |\hat{G}_E|\bm{r}' \rangle =-\sum_{k}\langle \bm{r} |k\rangle \langle k|\bm{r}' \rangle /(E_{k}-E)$ with $|k\rangle$ and $E_{k}$ being the eigenfunctions  and eigenvalues of a Hamiltonian, respectively. For the Hamiltonian (\ref{1dconfham}), $k=(n,m,q)$ is a multi-index, where $n=0,1,2,\ldots$ is the main quantum number, $m=0,\pm1,\pm2,\ldots$ is the orbital momentum in 2D, and $q$ is the $z$-component of the wavevector. The eigenvalues and normalized eigenfunctions are given by $E_{k}=\hbar\omega(2n+|m|)+\frac{q^2}{2m_{0}}$ and $\varphi_{k}(\rho,\alpha,z)=\frac{e^{iqz}}{\sqrt{2\pi}}\phi_{n,m}(\rho,\alpha)$, respectively. Here $\phi_{n,m}(\rho,\alpha)$ are the eigenvalues for the 2D oscillator in the polar coordinates $x=\rho\cos\alpha$, $y=\rho\sin\alpha$. One can show that for an arbitrary main quantum number, $\phi_{n,0}(\rho=0,\alpha)=1/(\sqrt{\pi}l_{0})$ and $\phi_{n,m}(\rho=0,\alpha)=0$ for $m\not=0$. Since we are interested at the GF of the Hamiltonian (\ref{1dconfham}) only at $\rho=\rho'=0$, we are left with one sum over the main quantum number and thus arrive at Eq.~(\ref{GFcyl}).

When the scattering momentum is sufficiently small, $p<2/l_{0}$, only the first term in the sum in Eq.~(\ref{GFcyl}) is singular. Separating this term and using relations $\int_{-\infty}^{\infty}\frac{dq}{2\pi}\frac{e^{iq z}}{(q^2+\varkappa^2)} =\frac{e^{-\varkappa|z|}}{2\varkappa}$ and $\text{p.v.}\!\!\int_{-\infty}^{\infty}\frac{dq}{2\pi}\frac{e^{iq z}}{(p^2-q^2)} =\frac{\sin p|z|}{2p}$ yield
\begin{align}
  \frac{\hbar^2}{2m_{0}}G_{p}(z)=&\frac{1}{\pi l_{0}^2}\left[\frac{\sin p|z|}{2p}-\frac{l_{0}}{4}\Lambda\left(\frac{2|z|}{l_{0}},-\frac{p^2l_{0}^2}{4}\right)\right], \label{GFcyl1}\\ \Lambda(\xi,\epsilon)=&\sum_{n=1}^{\infty}\frac{e^{-\xi\sqrt{n+\epsilon}}}{\sqrt{n+\epsilon}},\label{Lamdef}
\end{align}
where dimensionless variables $\xi =2|z|/l_{0}$ and $\epsilon =-p^2l_{0}^2/4$ are introduced.

The first term of the GF (\ref{GFcyl1}) is proportional to the one-dimensional GF of a free particle, which obeys the relation $\left(\frac{\partial^2}{\partial z^2}+p^2\right)\frac{\sin p|z|}{2p} =\delta(z)$.  At large distances $|z|\gg l_{0}$, the second term of the GF falls off exponentially as it follows directly from the expansion (\ref{Lamdef}). Then the asymptotics of the scattering part of the wavefunction, proportional to the GF, is essentially one-dimensional, as expected.

Let us calculate the GF at low momenta $|\epsilon|\ll 1$ and small coordinates $\xi\ll 1$. To expand $\Lambda(\xi,\epsilon)$ in $\xi$ and $\epsilon$ around zero, we use the identity for exponential $e^{-x}=\frac{1}{\sqrt{\pi}}\int_{0}^{\infty}\frac{dt}{t^{3/2}}e^{-1/t}e^{-x^2t/4}$ and find the integral representation
\begin{align}
 \Lambda(\xi,\epsilon)=\frac{1}{\sqrt{\pi}}\int_{0}^{\infty}\frac{dt}{t^{3/2}}e^{-\frac{1}{t}}e^{-\frac{\xi^2(1+\epsilon)t}{4}}
                        \Phi\left(e^{-\frac{\xi^2 t}{4}},\frac{1}{2},1+\epsilon\right),\label{intLam}
\end{align}
where $\Phi\left(x,s,\alpha\right)=\sum_{n=0}^{\infty}x^n/(n+\alpha)^{s}$ is the Lerch transcendent, or Lerch zeta function~\cite{erdelyiI}. Upon expanding the integrand
\begin{align}
e^{-\frac{\xi^2(1+\epsilon)t}{4}}\Phi\left(e^{-\frac{\xi^2t}{4}},\frac{1}{2},1+\epsilon\right)=&\frac{2 \sqrt{\pi }}{\sqrt{t} \xi} +\zeta({1}/{2}) -\frac{\epsilon}{2}\zeta({3}/{2}) \nonumber\\
& + O(\xi^2) + O(\epsilon^2)\nonumber
\end{align}
and substituting it into Eq.~(\ref{intLam}), we obtain
\begin{align}
\Lambda(\xi,\epsilon)=&\frac{2}{\xi}+\zeta({1}/{2})+\frac{\xi}{2}-\frac{\epsilon}{2}\zeta({3}/{2})+O(\xi^2)+ O(\epsilon^2)\nonumber \\
                      &+O(\xi\epsilon) \label{intLamexp}
\end{align}
with $\zeta(x)$ being the Riemann zeta function. The symbol $O(x)$ is the big $O$ notation, which designates terms of the order of $x$ or smaller. When $|z|\ll l_{0}$ and $p\ll 1/l_{0}$, the expansion of the GF (\ref{GFcyl1}) takes the form
\begin{align}
\frac{\hbar^2}{2m_{0}}G_{p}(z)=&-\frac{1}{4\pi|z|}\!-\!\frac{1}{4\pi l_{0}}\zeta({1}/{2})\!+\!\frac{|z|}{4\pi l_{0}^2}\!-\!\frac{p^2l_{0}}{32\pi}\zeta({3}/{2})\nonumber\\
&+O(z^2/l_{0}^{2})+O(|z|\, p^2 l_{0})+O(p^4 l_{0}^{4}).\label{GFcylfin}
\end{align}

\subsection{Flat confinement}
\label{sec:flat}

The flat harmonic confinement along the $z$-axis is considered by close analogy with the previous section. A particle, being confined in $z$-direction, can move freely in the $x$-$y$ plane. The Hamiltonian reads
\begin{align}\label{2dcondharm}
  \hat{H}=-\frac{\hslash^2\nabla^2}{2m_{0}} + \frac{m_{0}\omega^2z^2}{2}-E_0
\end{align}
with the shift $E_0=\hbar\omega/2$. A complete set of its eigenfunctions is given by $\varphi_{\bm{q},n}(\bm{\rho},z) =\phi_{n}(z)e^{i\bm{q}\cdot\bm{\rho}}/(2\pi)$ with $\phi_{n}(z)$ being the normalized eigenfunctions of the linear harmonic oscillator. We need only their absolute values at $z=0$: when $n$ is odd then $\phi_{n}(0)=0$, while for even $n$
\begin{align}\label{1Dosc0}
|\phi_{2k}(0)|=\frac{1}{\pi^{1/4} l_{0}^{1/2}}\frac{\sqrt{(2k)!}}{2^{k}k!},\quad k=0,1,2,\ldots,
\end{align}
see, e.g., the textbook~\cite{llvol3_77}.

Using the same method as in the previous section, we find the expression for the GF at $z=z'=0$ and $E=\frac{\hbar^2p^2}{2m_{0}}$
\begin{align}
  &G_{p}(|\bm{\rho}-\bm{\rho}'|)\equiv\langle\bm{\rho},z=0|\hat{G}_E|z'=0,\bm{\rho}'\rangle\nonumber \\
  &=-\sum_{k=0}^{\infty}|\phi_{2k}(0)|^2\text{p.v.}\!\!\int\frac{d^2q}{(2\pi)^2}\frac{e^{i\bm{q}\cdot(\bm{\rho}-\bm{\rho}')}}{2\hbar\omega k+\frac{\hbar^2}{2m_{0}}(q^2-p^2)},
\label{GFflat}
\end{align}
where $|\phi_{2k}(0)|$ is given by Eq.~(\ref{1Dosc0}). When the momentum is small, $p<2/l_{0}$, only the first term in the sum is singular. Separating it and using the identities $\text{p.v.}\!\!\int \frac{d^2q}{(2\pi)^2}\frac{e^{i\bm{q}\cdot \bm{\rho}}}{p^{2}-q^{2}} =\frac{Y_{0}(p\rho)}{4}$, $\int\frac{d^2q}{(2\pi)^2}\frac{e^{i\bm{q}\cdot \bm{\rho}}}{\varkappa^{2}+q^{2}} =\frac{K_{0}(\varkappa \rho)}{2\pi}$ for positive $p$ and $\varkappa$ yield
\begin{align}
\frac{\hbar^2}{2m_{0}}G_{p}(\rho)=&\frac{1}{\sqrt{\pi}l_{0}}\left[\frac{Y_{0}(p\rho)}{4}-\frac{1}{2\pi}S\left(\frac{2\rho}{l_{0}},-\frac{p^2l_{0}^2}{4}\right)\right], \label{GFflat1}\\
  S(\xi,\epsilon)=&\sum_{k=1}^{\infty}K_{0}(\xi\sqrt{k+\epsilon})\frac{(2k)!}{4^{k}(k!)^2} \label{Lamdef2D}
\end{align}
with the dimensionless variables $\xi=2\rho/l_{0}$ and $\epsilon=-p^2l_{0}^2/4$. Here $Y_{0}(x)$ and $K_{0}(x)$ are the Bessel and modified Bessel functions of the second kind, respectively (see, e.g., Ref.~\cite{abr_steg64}, Ch.~9).

Similar to the one-dimensional case considered in Sec.~\ref{sec:cyl}, the first term is proportional to the GF of a free particle in two dimensions: $\left(\frac{\partial^2}{\partial \rho^2}+\frac{1}{\rho}\frac{\partial}{\partial \rho}+p^2\right)\frac{Y_{0}(p\rho)}{4} =\delta(\bm{\rho})$. When $\rho\gg l_{0}$, the second term in Eq.~(\ref{GFflat1}) is exponentially small, which actually implies two-dimensional scattering.

The relation $K_{0}(x)=\int_{0}^{\infty}\frac{dt}{2t}e^{-t}e^{-x^2/(4t)}$ enables us to sum up the series (\ref{Lamdef2D}) under the integral and find its integral representation in elementary functions
\begin{align}
S(\xi,\epsilon)=\int_{0}^{\infty}\frac{dt}{2t}\exp\left(-t-\frac{\epsilon\,\xi^2}{4t}\right)\left[\frac{1}{\sqrt{1-e^{-\xi^2/(4t)}}}-1\right].\label{intS}
\end{align}
This relation is a particular case of a more general equation, which was obtained by another method in Refs.~\cite{Idziaszek06,Liang08}. This general equation is actually the Fourier transform of the well-known Feynman propagator for a harmonic oscillator.

It is easy to obtain the expansion
\begin{align}
  S(\xi,\epsilon)=&\frac{\sqrt{\pi}}{\xi}+\ln\xi +\gamma-\sqrt{\pi}-\ln 2+C-\xi\frac{\sqrt{\pi}}{8}-\epsilon \ln 2\nonumber\\
  &+O(\xi^2)+ O(\epsilon^2)+O(\xi\epsilon),\label{Sfin}
\end{align}
where $\gamma=0.5772\ldots$ is Euler's constant and
\begin{align}
C\!=\!\int_{0}^{\infty}\!\frac{dy\, e^{-y}}{2y}\left[\frac{1}{\sqrt{1-e^{-y)}}(1+\sqrt{1-e^{-y)}})}\!+\!1-\!\frac{1}{\sqrt{y}}\right].\nonumber
\end{align}
Its numerical value is given by $C=0.8035589\ldots$.

To find the GF at $\rho\ll l_{0}$ and $p\ll 1/l_{0}$, we substitute the expansions (\ref{Sfin}) and $Y_0(x)=2\ln(xe^{\gamma}/2)/\pi + O(x^2\ln x)$ into Eq.~(\ref{GFflat1}) and finally arrive at
\begin{align}
\frac{\hbar^2}{2m_{0}}G_{p}(\rho)=&-\frac{1}{4\pi\rho}+\frac{1}{4\pi^{3/2}l_{0}}\ln\left(\frac{\pi p^2l_{0}^2}{2B}\right)\nonumber\\
&+\frac{1}{8\pi l_{0}}\left[ \frac{\rho}{l_{0}}-\frac{\ln2}{\sqrt{\pi}}p^2l_{0}^2\right]+O(\rho^2 p^2\ln p\rho)\nonumber\\
&+O(\rho^2/l_{0}^2)+O(\rho p^2l_{0})+O(p^4l_{0}^4), \label{GFflatfin}
\end{align}
where we put by definition
\begin{align}
B\equiv2\pi\exp[2(C-\sqrt{\pi})]=0.904916\ldots. \label{Bconst}
\end{align}
The first two terms of GF (\ref{GFflatfin}) were first calculated in \cite{Petrov00} with the constant $B$ being equal to $1$. In the subsequent paper \cite{Petrov01}, its numerical value was estimated to be $0.915$. Further, the value of $B$ being equal to $0.9049$ was obtained in Ref.~\cite{Idziaszek06,*Idziaszek15}, which is consistent with Eq.~(\ref{Bconst}).
We believe that this is the most accurate estimation.

%------------------------------------------------------------
\section{Confinement-induced resonance in one dimension: cylindrical geometry}
\label{CIR1D}

\subsection{The tight confinement}
\label{TC1D}

When the bosons are trapped in the $x$-$y$ transverse direction with the axially symmetric harmonic potential $V_\mathrm{ext}(\rho) =m\omega^2 \rho^2/2$, the order parameter depends only on the transverse coordinates, but it is homogeneous along the $z$-axis.

The tight confinement implies that the chemical potential is small in scale of $\hbar\omega$, that is, $l\ll \xi_\mathrm{h}$, in terms of the oscillator length $l =\sqrt{\hbar/(m\omega)}$. Then with a good accuracy one can put for the wavefunction of BEC \cite{petrov00a} $\phi =\phi(\rho_1) =\sqrt{n_\mathrm{1D}}\phi_{0}(\rho_1)$ with $\phi_0(\rho_1) =\frac{\exp[-\rho^2_1/(2l^2)]} {l\sqrt{\pi}}$ being the ground state wavefunction of the one-particle Schr\"odinger equation with Hamiltonian (\ref{1bodyHam}), and $n_\mathrm{1D}$ is the linear density of the Bose-Einstein condensate. Thus, the nonlinear term in Eq.~(\ref{1body}) is treated as a perturbation, and the chemical potential is obtained by multiplying Eq.~(\ref{1body}) by $\phi_{0}(\rho_{1})$ and integrating out over $\bm{r}_{1}$:
\begin{align}\label{mu1D}
  \mu=\frac{1}{L}\iint d\bm{r}_{1}d\bm{r}_{2}\phi_{0}(\rho_{1})\phi_{0}(\rho_{2})V(|\bm{r}_{1}\!-\!\bm{r}_{2}|)\varphi(\bm{r}_{1},\bm{r}_{2}),
\end{align}
where $L$ is the length of the system along the $z$-axis.

To take the integral (\ref{mu1D}), we need to examine the properties of the pair wavefunction. In the case of transverse confinement, Eqs.~(\ref{1body}) and (\ref{2body}) are translationally invariant under a spatial translation $z_1\to z_1+b$, $z_2\to z_2+b$ for any $b$, which yields $\varphi =\varphi(\bm{\rho}_{1},\bm{\rho}_{2},|z_{1}-z_{2}|)$. Moreover, for the harmonic confinement, the spatial variables in Eq.~(\ref{2body}) are separated into the center-of-mass position and relative coordinates by analogy with the two-body problem. We can look for a solution in the form
\begin{align}\label{phi_rel_intro}
\varphi(\bm{r}_1,\bm{r}_2)=\varphi(\rho,z)\exp\left[-\big(\bm{\rho}_1+\bm{\rho}_2\big)^2/(4l^2)\right]
\end{align}
and the same relation for $\psi(\bm{\rho}_1,\bm{\rho}_2,z)$, see Eq.~(\ref{psidef}). Here $\bm{\rho}=\bm{\rho}_1-\bm{\rho}_2$ and $z=z_{1}-z_{2}$ are the relative coordinates, and $\rho =|\bm{\rho}|$. The boundary conditions (\ref{bc}) read $\varphi(\rho,z) \simeq n_{\rm 1D}\exp[-\rho^2/(4l^2)]/(\pi l^2)$ when $z\gg\xi_\mathrm{h}$.

In the range $r\ll l$, one can use the approximation
\begin{align}
\varphi(\rho,z)=\eta\varphi^{(0)}_{\rm 3D}(r)\frac{n_{\rm 1D}}{\pi l^2},
\label{phiapp}
\end{align}
since $r_0\ll l$. Here $r=\sqrt{\rho^2+z^{2}}$, and $\varphi^{(0)}_{\rm 3D}(r)$ is the 3D solution of the two-body Schr\"odinger equation in the center-of-mass system at zero energy
\begin{equation}
-(\hbar^2/m)\nabla^2\varphi^{(0)}_{\rm 3D}(r)+V(r)\varphi^{(0)}_{\rm 3D}(r)=0.\nonumber
\end{equation}
The solution is assumed to be regular in the vicinity of the origin and obeys the boundary condition when $r\gg r_0$
\begin{align}
\varphi^{(0)}_{\rm 3D}(r)\simeq&\, 1-a/r, \label{phi3dasymp}\\
a=&\,\frac{m}{4\pi\hbar^2}\int d\bm{r}V(r)\varphi^{(0)}_{\rm 3D}(r). \label{intVphi}
\end{align}
Substituting Eqs.~(\ref{phi_rel_intro}) and (\ref{phiapp}) into Eq.~(\ref{mu1D}) and using the difference of the  scales $r_0\ll l$ and Eq.~(\ref{intVphi}), we are left with
\begin{align}\label{mu1D1}
  \mu=\eta\frac{2\hbar^2a}{m\,l^2}n_{\rm 1D}.
\end{align}

Here the point is that the constant $\eta\not=1$, which is crucial for the appearance of CIR. This constant can be recovered from Eq.~(\ref{2body}).

\subsection{The application of the Green's function}
\label{AppGF1}

Substituting Eq.~(\ref{phi_rel_intro}) into Eq.~(\ref{2body}), we arrive at the equation for the wavefunction $\varphi(\rho,z)$:
\begin{align}\label{phicm}
\left[-\frac{\hbar^2\nabla^2}{2m_{0}}+\frac{m_{0}\omega^2\rho^2}{2}-E_{0}-2\mu\right]\psi(\rho,z)=-V(r)\varphi(\rho,z)
\end{align}
with $m_{0}=m/2$ being the reduced mass. It is sufficient to find the wavefunction $\varphi(\rho,z)$ at $\rho=0$. According to Eqs.~(\ref{phiapp}) and (\ref{phi3dasymp}), it has the asymptotics
\begin{align}\label{phiapp_short}
\varphi(\rho=0,z)\simeq \eta\frac{n_{\rm 1D}}{\pi l^2}\left(1-\frac{a}{|z|}\right)
\end{align}
in the short-range regime $r_{0}\ll |z|\ll l$.

The solution of Eq.~(\ref{phicm}) can be found with the Green's function (GF) at $E=2\mu=\frac{\hbar^2p^2}{2m_{0}}$ as $\psi(\rho,z) =\int d\bm{r}' \langle z,\bm{\rho}|\hat{G}_E|\bm{\rho}',z'\rangle V(r')\varphi(\rho',z')$. In the short-range regime $|z|\ll l$ and at low energies $p\ll 1/l$, the GF at $\rho=\rho'=0$ was obtained in Sec.~\ref{sec:cyl} and given by Eq.~(\ref{GFcylfin}). At $m_{0}=m/2$ and $l_{0}=\sqrt{\hbar/(m_0 \omega)}=\sqrt{2}l$, we get
\begin{align}
  \varphi(\rho=0,z)=\frac{n_{\rm 1D}}{\pi l^2}+G_{p}(z)\int d\bm{r}'V(r')\varphi(\rho',z'),\nonumber
\end{align}
where Eq.~(\ref{psidef}) is used: $\varphi(\rho=0,z) =n_{\rm 1D}/(\pi l^2) +\psi(\rho=0,z)$.

The potential $V(r)$ is localized within $r\lesssim r_{0}$, where the wavefunction $\varphi(\rho,z)$ takes the form (\ref{phiapp}). Using the relation (\ref{intVphi}) and the first two terms of Eq.~(\ref{GFcylfin}) for the Green's function, we find
\begin{align}
  \varphi(\rho=0,z)\simeq\frac{n_{\rm 1D}}{\pi l^2}\left[1-\eta\, a\left(\frac{1}{|z|}+\frac{\zeta(1/2)}{\sqrt{2}l}\right)\right].\nonumber
\end{align}
Comparing this asymptotics with Eq.~(\ref{phiapp_short}) finally yields
\begin{align}\label{etafin}
  \eta=\left(1+\frac{a\,\zeta(1/2)}{\sqrt{2}l}\right)^{-1},
\end{align}
where $\zeta(1/2)=-1.46035\ldots$.

Note that in our paper \cite{cherny04} we used the approximation for the wavefunction $\varphi(\rho,z)=\eta(1-a/r)\exp[-\rho^2/(4l^2)]{n_{\rm 1D}}/{(\pi l^2)}$ when $r_0\ll r \ll \xi_\mathrm{h}$. It allows us to integrate out the $\rho$ coordinate in Eq.~(\ref{phicm}) and obtain the coefficient $\eta$ without using the GF formalism. However, as is mentioned in the Introduction, this leads to an incorrect coefficient in the denominator of Eq.~(\ref{etafin}): $-\sqrt{\pi}=-1.772\ldots$ instead of the correct one $\zeta(1/2)=-1.46035\ldots$. The reason is that this approximation is accurate only when $r_0\ll r \ll l$ [see Eq.~(\ref{phiapp})] but fails near $r\simeq l$. Thus, the approximation of Ref.~\cite{cherny04} can be used for qualitative rather than quantitative estimations.

\subsection{The chemical potential}
\label{sec:mu1D}

The chemical potential is given by Eqs.~(\ref{mu1D1}) and (\ref{etafin})
\begin{align}
\mu&=g_{\text{1D}}n_{\rm 1D},\label{mu1D1fin}\\
g_{\text{1D}}&= \frac{2\hbar^2a}{m\,l^2}\frac{1}{1+\frac{a\,\zeta(1/2)}{\sqrt{2}l}}. \label{eff_g1D}
\end{align}
Taking into account the relation $g_{\text{1D}}=-\frac{2\hbar^2}{m\, a_\mathrm{1D}}$ for the coupling constant in 1D, we arrive at the expression for the effective 1D scattering length $a_\mathrm{1D}$ in the two-body problem (see, e.g., Ref.~\cite{jeszenszki18}), first obtained in Ref.~\cite{olshanii98}
\begin{align}
{a_\mathrm{1D}}=-\frac{l^2}{a}\left(1+\frac{a\,\zeta(1/2)}{\sqrt{2}l}\right).\nonumber
\end{align}

%------------------------------------------------------------
\section{Confinement-induced resonance in two dimensions: flat geometry}
\label{CIR2D}

\subsection{The tight confinement}
\label{TC2D}

The flat confinement of identical bosons is provided by harmonic forces along the $z$-axis: $V_\mathrm{ext}(z) =m\omega^2 z^2/2$. The order parameter is homogeneous within the $x$-$y$ plane, and in the case of the tight confinement $l\ll \xi_\mathrm{h}$, one can put with a good accuracy $\phi =\phi(z_1) =\sqrt{n_\mathrm{2D}}\phi_{0}(z_1)$ with the oscillator ground-state wavefunction $\phi_0(z_1) =\exp[-z_1^2/(2l^2)] /(l^{1/2}\pi^{1/4})$. Here $n_\mathrm{2D}$ is the 2D density of the Bose-Einstein condensate. The chemical potential can be obtained from Eq.~(\ref{1body}) by analogy with the previous section
\begin{align}\label{mu2D}
  \mu=\frac{1}{S}\iint d\bm{r}_{1}d\bm{r}_{2}\phi_{0}(z_{1})\phi_{0}(z_{2})V(|\bm{r}_{1}\!-\!\bm{r}_{2}|)\varphi(\bm{r}_{1},\bm{r}_{2}),
\end{align}
where $S$ is the area of the system occupied in the $x$-$y$-plane.

The pair wavefunction is invariant under translations along the $x$-$y$ plane: $\varphi(\bm{r}_1,\bm{r}_2)=\varphi(z_{1},z_{2},\rho)$. It can be written for the harmonic confinement as $\varphi(\bm{r}_1,\bm{r}_2)=\varphi(z,\rho)\exp\left[-(z_1+z_2)^2/(4l^2)\right]$.

The approximation for the pair wavefunction for $r\ll l$ is
\begin{align}
\varphi(z,\rho)=\eta\varphi^{(0)}_{\rm 3D}(r)\frac{n_{\rm 2D}}{l\sqrt{\pi}},
\label{phiapp2D}
\end{align}
which is analogous to Eq.~(\ref{phiapp}). In the same manner, we find from Eqs.~(\ref{intVphi}), (\ref{mu2D}) and (\ref{phiapp2D})
\begin{align}
  \mu=\frac{4\pi\hbar^2}{m}n_{\rm 2D} u, \label{mu2D1}
\end{align}
where we introduce a more convenient dimensionless parameter
\begin{align}\label{udef}
u=\frac{\eta a}{\sqrt{2\pi}l}.
\end{align}

The prefactor $\eta$ can also be found by analogy with Sec.~\ref{CIR1D}.

\subsection{The application of the Green's function}
\label{AppGF2}

On the one hand, in the short-range regime $r_{0}\ll \rho \ll l$, Eqs.~(\ref{phi3dasymp}) and (\ref{phiapp2D}) yield
\begin{align}\label{phiapp_short2D}
\varphi(z=0,\rho)\simeq \eta\frac{n_{\rm 2D}}{l\sqrt{\pi}}\left(1-\frac{a}{\rho}\right).
\end{align}
One the other hand, one can use the GF for Eq.~(\ref{2body}) (see Sec.~\ref{sec:flat}) and obtain in the same regime
\begin{align}
  \varphi(z=0,\rho)\simeq&\,\frac{n_{\rm 2D}}{\sqrt{\pi} l}+G_{p}(\rho)\int d\bm{r}'V(r')\varphi(z',\rho')\nonumber\\
                        =&\,\frac{n_{\rm 2D}}{\sqrt{\pi} l}+G_{p}(\rho)\eta\frac{n_{\rm 2D}}{l\sqrt{\pi}}\frac{4\pi\hbar^2a}{m}
\label{psi_short2D}
\end{align}
with the GF (\ref{GFflatfin}), where $2\mu=\frac{\hbar^2p^2}{2m_{0}}$, $m_{0}=m/2$, and $l_{0}=\sqrt{2}l=\sqrt{2\hbar/(m \omega)}$.

\subsection{The chemical potential}
\label{sec:mu2D}

In the quasi-1D condensate, considered in the previous section \ref{CIR2D}, the GF remains finite in the limit $p\to 0$. By contrast, in the quasi-2D condensate, the GF (\ref{GFflatfin}) exhibits a logarithmic divergence as $p\to 0$. This implies that the chemical potential should be determined self-consistently. Indeed, comparing Eq.~(\ref{phiapp_short2D}) with Eq.~(\ref{psi_short2D}) and using the definition (\ref{udef}) yield the algebraic equation for $u$
\begin{align}
  &\frac{1}{u}+\ln u=\delta, \label{uself}\\
  &\delta=\frac{\sqrt{2\pi}l}{a}-\ln\left[\frac{8\pi^2 n_{\rm 2D}l^2}{B}\right], \label{delta}
\end{align}
where the constant $B$ is defined by Eq.~(\ref{Bconst}). Thus, the chemical potential is given by Eqs.~(\ref{mu2D1}), (\ref{uself}), and (\ref{delta}).

\begin{figure}[!tb]
\centerline{\includegraphics[width=.8\columnwidth]{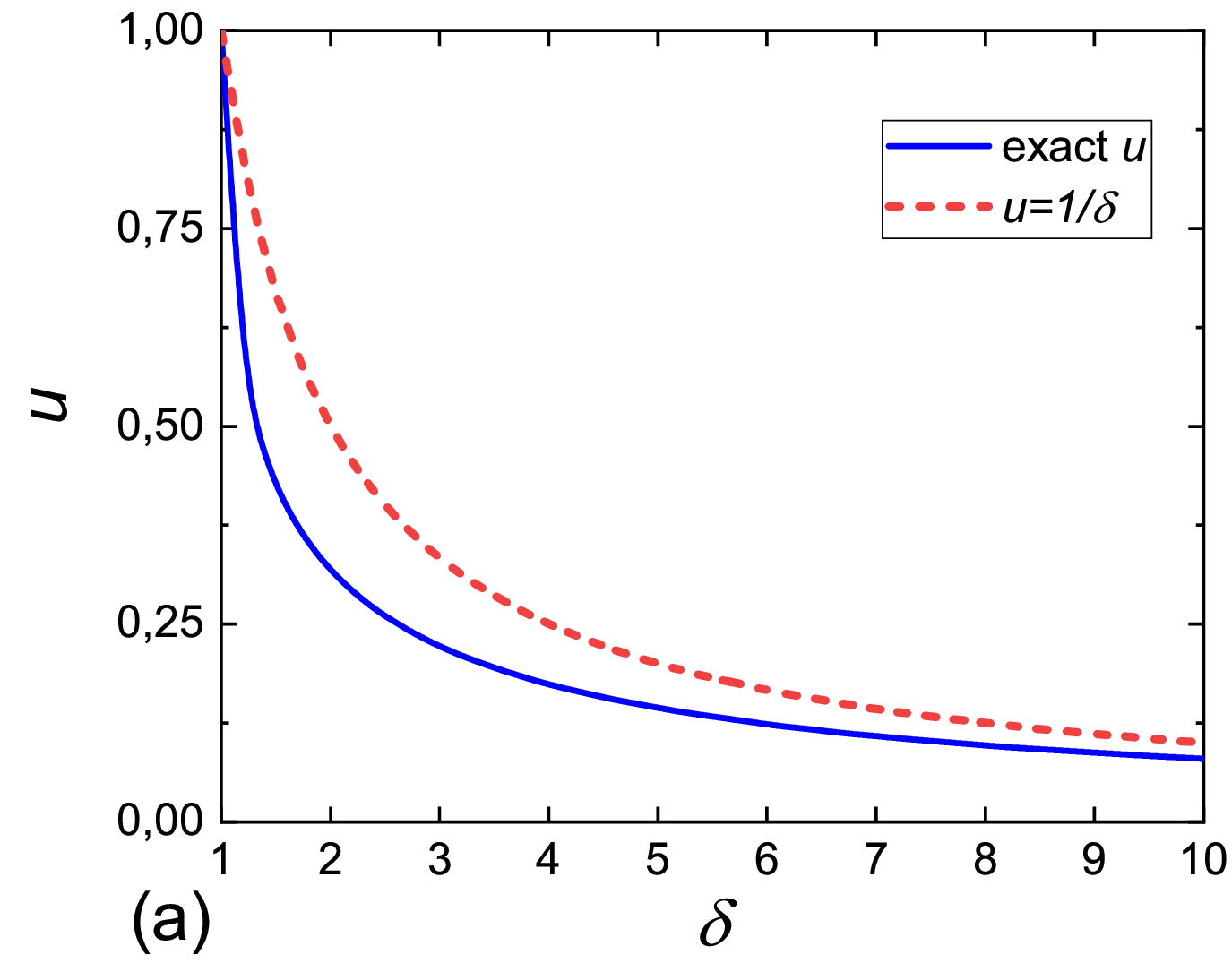}}
\centerline{\includegraphics[width=.8\columnwidth]{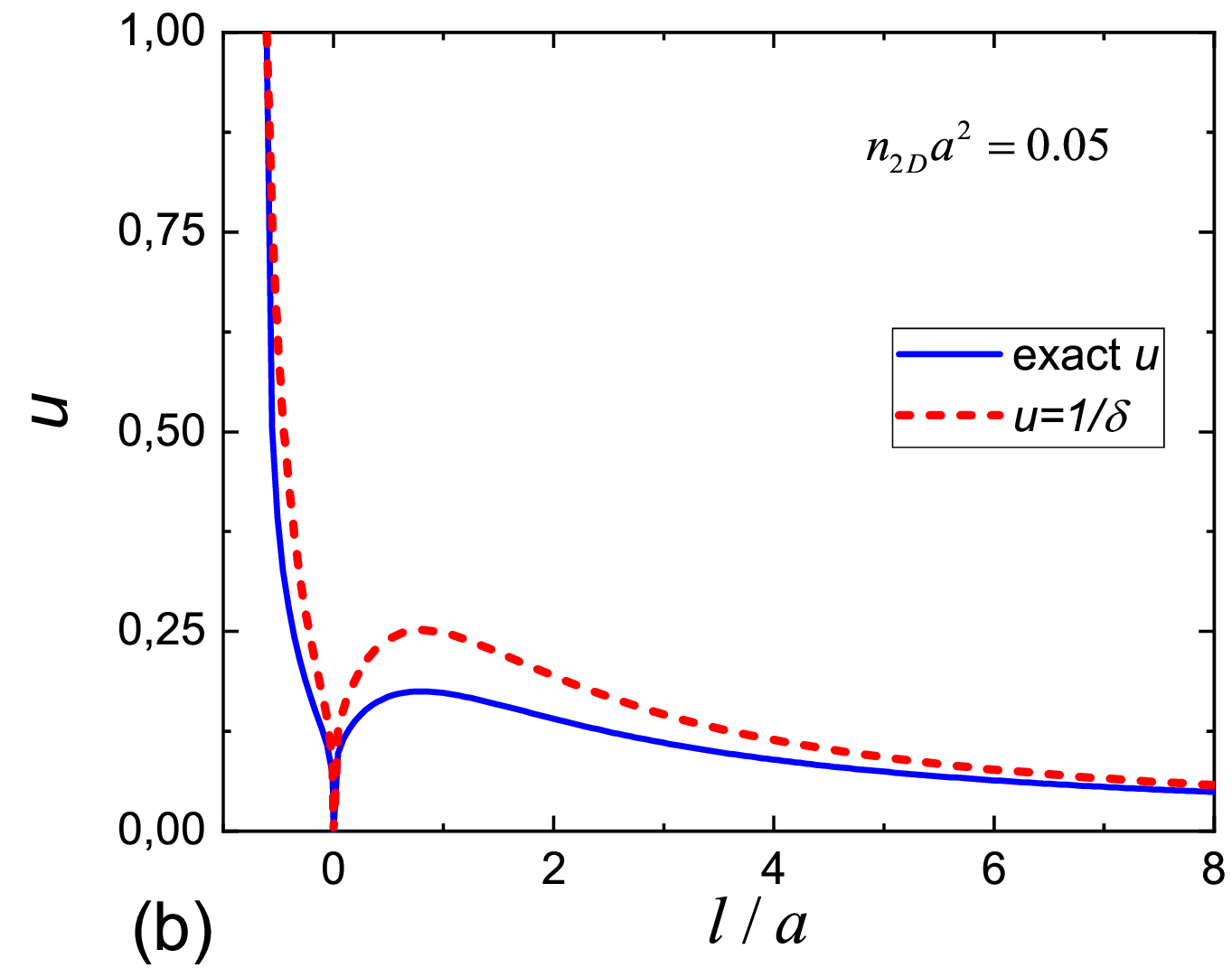}}
\caption{\label{fig:udel} (a) The solution of Eq.~(\ref{uself}), $u(\delta)$ (blue solid line). The parameter $u$ is the chemical potential in units of $4\pi\hbar^2 n_{\rm 2D}/m$, see Eq.~(\ref{mu2D1}). At large $\delta$, the function $u(\delta)$ asymptotically approaches $1/\delta$ (dashed red line). (b) The chemical potential in the pancake geometry [in units of $4\pi\hbar^2 n_{\rm 2D}/m$], obtained from Eqs.~(\ref{uself}) and (\ref{delta}), vs. the oscillator length $l$ of the trapping potential [in units of the 3D scattering length $a$] (blue solid line). The same diagram in the approximation $u\simeq1/\delta$ is shown in dashed red line. This approximation overestimates the values of the chemical potential. The absolute value of $a$ and 2D density $n_{\rm 2D}$ are supposed to be fixed, while the sign of $a$ can change. When $a$ is negative, the CIR is observed: the chemical potential grows rapidly with increase of the oscillator length, because the parameter $\delta$ [Eq.~(\ref{delta})] decreases. When the scattering length is positive, one can also observe a kind of smoothed resonance behaviour of the chemical potential with the maximum at $l/a=0.7978\ldots$, see Eq.~(\ref{la_ratio}). The position of the maximum is independent of the 2D density.
}
\end{figure}

Equation (\ref{uself}) has no real solutions when $\delta<1$ and two solutions when $\delta>1$: one is greater than 1, and the other is less than 1.  It follows from the thermodynamic stability condition $\frac{\partial \mu}{\partial n_{\rm 2D}} =\frac{4\pi\hbar^2}{m}\frac{u}{1-u}>0$ that only the solution $u<1$ is physical. This fits well with the condition of weak coupling, which tells us that the healing length should be at least smaller than the mean distance between bosons on the plane: $\frac{\hbar}{\sqrt{\mu m}}>\frac{1}{\sqrt{n_{\rm 2D}}}$, see Eq.~(\ref{mu2D1}). When the dimensionless parameter $\delta$ is big then $u\simeq 1/\delta$ and we arrive at the weak-coupling relation for the chemical potential
\begin{align}
\mu\simeq \frac{4\pi\hbar^2 n_{\rm 2D}}{m\delta}, \label{mu2Dweak}
\end{align}
which allows us to interpret $\delta$, given by Eq.~(\ref{delta}), as the resonance denominator.  Note, however, that this approximation overestimates the values of the chemical potential, see Fig.~\ref{fig:udel}a.

The CIR arises (see Fig.~\ref{fig:udel}b) when the 3D scattering length is negative while the 2D density is sufficiently small for $\delta$ to be positive and close to 1, which is its minimum possible value within our approach. At positive scattering length, a reminiscence of the resonance is also observed, when the dimensionless chemical potential reaches the local maximum at
\begin{align}
\frac{l}{a} =\sqrt{\frac{2}{\pi}}=0.7978\ldots   \label{la_ratio}
\end{align}
as it follows from Eq.~(\ref{delta}). This differs from the experimentally found \cite{haller10} value ${l}/{a}=1.19(3)$, at which the CIR is realized at positive scattering lengths.

Relations (\ref{mu2D1}), (\ref{uself}), and (\ref{delta}) should be used with some caveats. First, the quasi 2D regime imposes the restriction on the oscillator length: the characteristic wavevector of bosons $p\simeq \sqrt{n_{\rm 2D}}$ should be smaller than $2/l$ as discussed in Sec.~\ref{sec:flat}. This yields $\frac{l}{a}\lesssim \frac{2}{\sqrt{n_{\rm 2D}a^2}}$, which for the chosen parameter $n_{\rm 2D}a^2=0.05$ in Fig.~\ref{fig:udel}b is of order of 9. Second, we assume that the range of interaction potential is much smaller than the oscillator length: $r_{0}\ll l$. Thus, in the vicinity of $l=0$, the approximation of zero-range interactions ($r_{0}=0$) is used. We believe that the extremely tight confinement $r_{0}\gtrsim l$ also can be described by means of the GGP equations with methods similar to those considered in Ref.~\cite{naidon07}.

%------------------------------------------------------------
\section{Conclsusions}
\label{sec:concl}

Using the GGP equations (\ref{1body}) and (\ref{2body}), we obtained the correct resonance denominators, which determine the positions of CIRs in the cigar-shaped and pancake geometries, see Eqs.~(\ref{mu1D1fin}), (\ref{eff_g1D}), and (\ref{delta}), (\ref{mu2Dweak}).

The GGP equations, being a particular case of the HFB approximation (see Appendix \ref{sec:HFB}), are applicable for a weakly interacting Bose gas, strictly speaking. Nevertheless, the GGP equations correctly reproduce the effective coupling in the low-dimensional geometries. The reason is that these equations describe well the short-range correlations even inside the radius of interaction potential. This confirms that the GGP equations can be considered as a universal tool for describing BECs in various regimes. For instance, the method has good prospects for studying the extremely tight confinements, crossover between 3D and 2D or 1D regimes, and the dipolar BEC in various regimes, including quantum droplets \cite{Bottcher21}.

In addition, the integral representation for GF was found for the cylindrical confinement [Eqs.~(\ref{GFcyl1}), (\ref{intLam})], which is valid for small scattering wavevectors $p<2/l_{0}$ and arbitrary values of coordinates. It is convenient for obtaining the low-density expansion (\ref{GFcylfin}) of the GF.

Within our approach, we found the position (\ref{la_ratio}) of the smoothed resonance at positive scattering length in the pancake geometry. The position is independent of the density. This is nearly two-thirds of the experimentally observed value \cite{haller10}. The discrepancy might be attributed to condensate depletion at finite temperatures or to many-body effects that are beyond the used weak-coupling approximation.

In this paper, one species of bosons is considered. As a prospect, one can apply the GGP equations to study the CIR resonances for two species of bosons as well. To tackle the problem, we need to introduce two additional indices and solve a system of five coupled equations, which contain two corresponding chemical potentials. The task is quite complicated, since the center of mass and relative motions are not separated anymore due to different lengths of transverse confinement for different species. Green's function methods of the two-body scattering in the presence of transverse confinement \cite{Peano05,Kim06,*Melezhik07} can be expected to apply here as well. Note that for the cylindrical geometry, the $p$-wave  contribution to the scattering amplitude of distinguishable atoms turns out to be quite significant \cite{Kim06,*Melezhik07}.

\acknowledgments

The author is grateful to Azis Embareck for useful and fruitful discussions and Vladimir Melezhik for careful reading of the manuscript and useful comments.

\appendix

\section{Hartree-Fock-Bogoliubov approximation}
\label{sec:HFB}

Let make some remarks about the Hartree-Fock-Bogoliubov (HFB) method. We consider only the stationary HFB theory at zero temperature. The HFB approximation for fermions was suggested by Bogoliubov \cite{bogoliubov59} as a variational scheme formulated in terms of the normal and anomalous averages of two Fermi-field operators. These scheme can be simplified \cite{deGennes64,deGennes_book66,baranger63} if we note that these averages are given by bilinear forms of the $u$ and $v$ parameters of the Bogoliubov transformation. Then the variational procedure can be reformulated directly in terms of $u$ and $v$ parameters, and the resulting equations are called Bogoliubov-de Gennes equations.

For bosons, the relations analogous to the Bogoliubov-de Gennes equations can be written down \cite{pitaevskii61,Fetter72}. They are called the Bogoliubov equations. In addition to the $u$ and $v$ parameters, these equations contain the wavefunction of the inhomogeneous Bose-Einstein condensate.

On the other hand, one can also formulate the HFB method for bosons as a variational scheme with three variational functions: the condensate wavefunction and normal and anomalous averages of two Bose-field operators. Below we follow this scheme \cite{cherny04} and derive the generalized GP equations by further simplifying the HFB scheme.

\subsection{Hartree-Fock-Bogoliubov approximation for bosons in the variational formulation}
\label{sec:HFBvar}
The ground-state energy and mean number of interacting bosons in an external field is given by (see, e.g., textbooks \cite{feynman1972,bogbog82})
\begin{align}
E =&-\frac{\hslash^2}{2m}\int d x\left.\nabla^2_x\langle{\hat\Psi}^\dag(x'){\hat\Psi}(x)\rangle\right|_{x'=x}\nonumber \\
&+\int d x\, V_{\rm ext}(x)\langle{\hat\Psi}^\dag(x){\hat\Psi}(x)\rangle\nonumber \\
&+\frac{1}{2}\int\! d {x}_1 d {x}_2\,V({x}_1,{x}_2) \langle{\hat\Psi}^{ \dagger}({x}_1){\hat\Psi}^{\dagger}({x}_{2})
{\hat\Psi}({x}_{2}){\hat\Psi}({x}_1)\rangle,
\label{efull}\\
N=&\int d x\,\langle{\hat\Psi}^\dag(x){\hat\Psi}(x)\rangle,
\label{Nint}
\end{align}
respectively. Here $x=({\bf r},\sigma)$ are the coordinate and spin of a particle, respectively, and $\int d x\cdots = \sum_{\sigma}\int d{\bf r}\cdots$. The brackets $\langle\cdots\rangle$ stand for the ground-state average. The Bose field operators are denoted as ${\hat\Psi}(x)$ and ${\hat\Psi}^{ \dagger}(x)$.

The Bose-Einstein condensation implies that the global gauge symmetry is broken. Then one can separate the condensate contribution to the field operators: $\hat{\Psi}(x)=\phi(x) +\hat{\vartheta}(x)$ and $\hat{\Psi}^{\dag}(x)=\phi^{*}(x) +\hat{\vartheta}^{\dag}(x)$. Here $\phi(x)$ and $\hat{\vartheta}(x)$ are the numerical and operator parts, respectively, for which we have $\phi(x)=\langle\hat{\Psi}(x)\rangle$ and $\langle\hat{\vartheta}(x)\rangle=0$. Then the two-boson correlators read
\begin{align}
\langle{\hat\Psi}^\dag(x){\hat\Psi}(x')\rangle =\,& \phi^*(x)\phi(x') +\langle\hat{\vartheta}^\dag(x) \hat{\vartheta}(x')\rangle,\label{2bnorm}\\
\langle{\hat\Psi}(x){\hat\Psi}(x')\rangle =\,& \phi(x)\phi(x') +\langle\hat{\vartheta}(x) \hat{\vartheta}(x')\rangle. \label{2banom}
\end{align}
The normal two-boson correlator (\ref{2bnorm}) is nothing else but the one-body density matrix, and the anomalous correlator (\ref{2banom}) can be treated as the two-body wavefunction of the Bose-Einsten condensate \cite{cherny00pair,cherny00}.

The diagonal part of the two-body density matrix, which determines the interaction term, can also be found by separating the condensate part and using Wick's theorem for the correlators of $\hat{\vartheta}$ and $\hat{\vartheta}^\dag$ operators
\begin{align}
&\langle{\hat\Psi}^{\dagger}({x}_1){\hat\Psi}^{\dagger}({x}_{2}){\hat\Psi}({x}_{2}){\hat\Psi}({x}_1)\rangle
=\phi^*(x_1)\phi(x_2)^*\phi(x_2)\phi(x_1)\nonumber\\
&+\phi(x_1)\phi(x_2)\langle\hat{\vartheta}^\dag(x_1) \hat{\vartheta}^\dag(x_2)\rangle
+\phi^*(x_1)\phi^*(x_2)\langle\hat{\vartheta}(x_2) \hat{\vartheta}(x_1)\rangle\nonumber\\
&+\phi(x_1)\phi^*(x_2)\langle\hat{\vartheta}^\dag(x_1) \hat{\vartheta}(x_2)\rangle
+\phi^*(x_1)\phi(x_2)\langle\hat{\vartheta}^\dag(x_2) \hat{\vartheta}(x_1)\rangle\nonumber\\
&+\langle\hat{\vartheta}^\dag(x_1) \hat{\vartheta}^\dag(x_2)\rangle\langle\hat{\vartheta}(x_2) \hat{\vartheta}(x_1)\rangle\nonumber\\
&+\langle\hat{\vartheta}^\dag(x_1) \hat{\vartheta}(x_1)\rangle\langle\hat{\vartheta}^\dag(x_2) \hat{\vartheta}(x_2)\rangle\nonumber\\
&+\langle\hat{\vartheta}^\dag(x_1) \hat{\vartheta}(x_2)\rangle\langle\hat{\vartheta}^\dag(x_2) \hat{\vartheta}(x_1)\rangle.
\label{4bose}
\end{align}
Wick's theorem is applicable, since the effective HFB Hamiltonian contains only a bilinear form of creation and annihilation operators \cite{Fetter72}.

The normal $\langle\hat{\vartheta}^\dag(x) \hat{\vartheta}(x')\rangle$ and anomalous $\langle\hat{\vartheta}(x) \hat{\vartheta}(x')\rangle$ averages are not independent quantities. The relations between them at zero temperature can be obtained \cite{cherny04} using the fact that the HFB ground state is the quasiparticle vacuum. The relations do not depend explicitly on parameters of the Hamiltonian and take the form
\begin{align}
&\int d x\, \langle\hat{\vartheta}^\dag(x_1)\hat{\vartheta}^\dag(x)\rangle\langle\hat{\vartheta}(x)
\hat{\vartheta}(x_2)\rangle =\langle\hat{\vartheta}^\dag(x_1)\hat{\vartheta}(x_2)\rangle
\nonumber \\
&\phantom{\int d x\, \langle\hat{\vartheta}^\dag}
+\int d x\, \langle\hat{\vartheta}^\dag(x_1)\hat{\vartheta}(x)\rangle
\langle\hat{\vartheta}^\dag(x)\hat{\vartheta}(x_2)\rangle,
\label{fphicord1}
\end{align}
\begin{align}
&\int d x\, \langle\hat{\vartheta}^\dag(x)\hat{\vartheta}(x_1)\rangle\langle\hat{\vartheta}(x)\hat{\vartheta}(x_2)\rangle
=\int d x\, \langle\hat{\vartheta}(x_1)\hat{\vartheta}(x)\rangle
\nonumber\\
&\phantom{\int d x\, \langle\hat{\vartheta}^\dag(x_1)\hat{\vartheta}(x)\rangle\langle\hat{\vartheta}(x)}
\times\langle\hat{\vartheta}^\dag(x)\hat{\vartheta}(x_2)\rangle.
\label{fphicord}
\end{align}

Finally, the stationary HFB equations can be obtained by means of variation of $E-\mu N$, given by Eqs.~(\ref{efull})-(\ref{4bose}), under the conditions (\ref{fphicord1}) and (\ref{fphicord}). The variational functions are $\phi(x)$, $\phi^*(x)$, $\langle\hat{\vartheta}(x)\hat{\vartheta}(x')\rangle$, $\langle\hat{\vartheta}^\dag(x)\hat{\vartheta}^\dag(x')\rangle$, and $\langle\hat{\vartheta}^\dag(x)\hat{\vartheta}(x')\rangle$. We do not explicitly write the resulting equations.

The full HFB scheme is rather cumbersome, and additional simplifications of Eq.~(\ref{4bose}) are often used \cite{griffin96}. The simplest approximation is to neglect all the quantum correlators and leave only the wavefunction of the condensate. Then we arrive at the well-known GP functional with the help of the pseudopotential.

\subsection{Generalized Gross-Pitaevskii equations as a simplified version of the Hartee-Fock-Bogoliubov approximation}
\label{sec:GGP_HFB}

The GGP equations can be obtained \cite{cherny04} from the full HFB theory when the normal two-boson terms are completely neglected in Eq.~(\ref{4bose})
\begin{align}
&\langle{\hat\Psi}^{\dagger}({x}_1){\hat\Psi}^{\dagger}({x}_{2}){\hat\Psi}({x}_{2}){\hat\Psi}({x}_1)\rangle
\simeq\phi^*(x_1)\phi(x_2)^*\phi(x_2)\phi(x_1)\nonumber\\
&+\phi(x_1)\phi(x_2)\langle\hat{\vartheta}^\dag(x_1) \hat{\vartheta}^\dag(x_2)\rangle
+\phi^*(x_1)\phi^*(x_2)\langle\hat{\vartheta}(x_2) \hat{\vartheta}(x_1)\rangle\nonumber\\
&+\langle\hat{\vartheta}^\dag(x_1) \hat{\vartheta}^\dag(x_2)\rangle\langle\hat{\vartheta}(x_2) \hat{\vartheta}(x_1)\rangle.\nonumber
\end{align}
Besides, if the condensate depletion is small, one can neglect the second term in the r.h.s of Eq.~(\ref{fphicord1}), which is of the next order of the condensate depletion. Thus, we are left with the relation
\begin{align}\label{thetaapp}
\langle\hat{\vartheta}^\dag(x_1)\hat{\vartheta}(x_2)\rangle
\simeq\int d x\, \langle\hat{\vartheta}^\dag(x_1)\hat{\vartheta}^\dag(x)\rangle
\langle\hat{\vartheta}(x)\hat{\vartheta}(x_2)\rangle.
\end{align}
In the approximation (\ref{thetaapp}), Eq.~(\ref{fphicord}) turns into identity.

Then the variational procedure yields \cite{cherny04} the GGP equations (\ref{1body})-(\ref{1bodyHam}).

\bibliography{GGP}

\end{document}